 \documentclass{aa}
\usepackage{graphicx}
\usepackage{txfonts}
\usepackage{natbib}
\begin{document}

%\draft
\def\be{\begin{equation}}
\def\ee{\end{equation}}
\def\bea{\begin{eqnarray}}
\def\eea{\end{eqnarray}}
\def\c{\cite}
\def\nn{\nonumber}
\def\mcr{{{\rm M_{cr}}}}
\def\xo{{X_{0}}}
\def\dm{\Delta {\rm M}}
\def\ms{\ifmmode {\rm M}_{\odot}{ ~ } \else ${\rm M}_{\odot}{ ~ }$ { }\fi}
\def\mb{m_{B}}
\def\bo{B_{o}}
\def\cxo{C(x_{o})}
\def\rsix{R_{6}}
\def\rnin{R_{9}}
\def\vr{v_{{\rm r}}}
\def\vro{v_{{\rm ro}}}
\def\vt{v_{\theta}}

\def\et{ {\it et al.}}
\def\la{ \langle}
\def\ra{ \rangle}
\def\ov{ \over}
\def\ep{ \epsilon}

\def\ep{\epsilon}
\def\th{\theta}
\def\ga{\gamma}
\def\Ga{\Gamma}
\def\la{\lambda}
\def\si{\sigma}
\def\al{\alpha}
\def\pa{\partial}
\def\de{\delta}
\def\De{\Delta}
\def\rsr{{r_{s}\over r}}
\def\rmo{{\rm R_{M0}}}
\def\rrm{{R_{{\rm M}}}}
\def\rra{{R_{{\rm A}}}}

\def\mdot{\ifmmode \dot M \else $\dot M$\fi}    % accretion rate
\def\mxd{\ifmmode \dot {M}_{x} \else $\dot {M}_{x}$\fi}
\def\med{\ifmmode \dot {M}_{Edd} \else $\dot {M}_{Edd}$\fi}
\def\bff{\ifmmode B_{{\rm f}} \else $B_{{\rm f}}$\fi}

\def\apj{\ifmmode ApJ \else ApJ \fi}    % lower
\def\apjl{\ifmmode  ApJ \else ApJ \fi}    %
\def\apjs{\ifmmode  ApJS \else ApJS \fi}
\def\aap{\ifmmode A\&A \else A\&A\fi}
\def\aaps{\ifmmode A\&AS \else A\&AS\fi}    %
\def\mnras{\ifmmode MNRAS \else MNRAS \fi}    %
\def\nat{\ifmmode Nature \else Nature \fi}
\def\prl{\ifmmode Phys. Rev. Lett. \else Phys. Rev. Lett.\fi}
\def\prd{\ifmmode Phys. Rev. D. \else Phys. Rev. D.\fi}
\def\pasp{\ifmmode  PASP \else PASP \fi}

\def\ms{\ifmmode {\rm M}_{\odot} \else ${\rm M}_{\odot}$\fi}    % lower
\def\na{\ifmmode \nu_{A} \else $\nu_{A}$\fi}    % Alfven frequency
\def\nk{\ifmmode \nu_{K} \else $\nu_{K}$\fi}    % Keplerian frequency
\def\ns{\ifmmode \nu_{{\rm s}} \else $\nu_{{\rm s}}$\fi}
\def\no{\ifmmode \nu_{1} \else $\nu_{1}$\fi}    % lower
\def\nt{\ifmmode \nu_{2} \else $\nu_{2}$\fi}    % upper
\def\ntk{\ifmmode \nu_{2k} \else $\nu_{2k}$\fi}    % upper
\def\dnmax{\ifmmode \Delta \nu_{max} \else $\Delta \nu_{2max}$\fi}
\def\ntmax{\ifmmode \nu_{2max} \else $\nu_{2max}$\fi}    % upper
\def\nomax{\ifmmode \nu_{1max} \else $\nu_{1max}$\fi}    % upper
\def\nh{\ifmmode \nu_{\rm HBO} \else $\nu_{\rm HBO}$\fi}    % HBO
\def\nqpo{\ifmmode \nu_{QPO} \else $\nu_{QPO}$\fi}    % HBO
\def\nz{\ifmmode \nu_{o} \else $\nu_{o}$\fi}    % HBO
\def\nht{\ifmmode \nu_{H2} \else $\nu_{H2}$\fi}    % HBO
\def\ns{\ifmmode \nu_{s} \else $\nu_{s}$\fi}    % stellar
\def\nb{\ifmmode \nu_{{\rm burst}} \else $\nu_{{\rm burst}}$\fi}
\def\nkm{\ifmmode \nu_{km} \else $\nu_{km}$\fi}    % stellar
\def\ka{\ifmmode \kappa \else \kappa\fi}    % stellar
\def\dn{\ifmmode \Delta\nu \else \Delta\nu\fi}
\def\dm{\ifmmode \Delta{}M \else \Delta{}M\fi}
\def\mdotsix{\ifmmode\dot{M}_{16} \else \dot{M}_{16}\fi}
\def\ps{\ifmmode P_{spin} \else P_{spin} \fi}

\def\sax{\ifmmode SAX J 1808.4-3658 \else SAX J 1808.4-3658\fi}

 \def\pspin{\ifmmode P_{s} \else $P_{s}$\fi}

\renewcommand{\vec}[1]{\mbox{\boldmath $\displaystyle #1$}}
\newcommand{\grad}{\vec{\nabla}}
\newcommand{\lap}{\nabla^2}
\newcommand{\vdot}{\vec{\cdot}}
\newcommand{\vcross}{\vec{\times}}
\newcommand{\divr}{\grad\vdot\,}
\newcommand{\curl}{\grad\vcross\,}
\newcommand{\avZ}{\left<Z\right>}
\def\rhof{\ifmmode \rho_{5} \else \rho_{5}\fi}
\def\rhos{\ifmmode \rho_{6} \else \rho_{6}\fi}
\def\mdotcr{\ifmmode \dot{M}_{cr} \else  \dot{M}_{cr}\fi}
\def\tohm{\ifmmode t_{ohmic} \else  t_{ohmic} \fi}

\def\tdif{\ifmmode t_{diff} \else  t_{diff} \fi}
\def\tacc{\ifmmode t_{accr} \else  t_{accr} \fi}
\newcommand{\gsimeq}{\mbox{$\, \stackrel{\scriptstyle >}{\scriptstyle
\sim}\,$}}
\newcommand{\lsimeq}{\mbox{$\, \stackrel{\scriptstyle <}{\scriptstyle
\sim}\,$}}

\title{Study of  measured pulsar masses and their possible conclusions}

\author{C.M. Zhang
        \inst{1}
        \and
        J. Wang
       \inst{1}
       \and
      Y.H. Zhao
        \inst{1}
       \and
      H.X. Yin
      \inst{2}
     \and
      L.M. Song
    \inst{3}
    \and
    D. P. Menezes
   \inst{4}\\
   % \and
    D. T.  Wickramasinghe
 \inst{5}
 \and
  L. Ferrario
 \inst{5}
 \and
 P.  Chardonnet
 \inst{6}
 }
\institute{National Astronomical Observatories,
 Chinese Academy of Sciences, Beijing 100012, China,\\
 \email{zhangcm@bao.ac.cn} \and
School of Space Science and Physics, Shandong University at Weihai,
Weihai 264209, China\\ \and
 Astronomical Institute, Institute of High Energy Physics,
 Chinese Academy of Sciences, Beijing 100039, China\\ \and
 Department  de F\'{\i}sica, CFM, Universidade Federal de
Santa  Catarina  Florian\'opolis, SC, CP. 476, CEP 88.040-900,
Brazil\\ \and Mathematical Sciences Institute, The Australian
National University, Canberra ACT 0200, Australia\\ \and  Lapth -
Lapp, Universit谷 de Savoie, B.P. 110, 74941 -Annecy-le-Vieux Cedex,
France }

\abstract { We study the statistics  of 61 measured masses of
neutron stars (NSs) in binary pulsar systems, including 18 double
NS (DNS) systems, 26 radio pulsars (10  in our Galaxy) with white
dwarf (WD) companions, 3 NSs with main-sequence companions, 13 NSs
in X-ray binaries, and one undetermined system. We derive a  mean
value of   $M = 1.46 \pm 0.30 \ms$. When the 46 NSs with measured
spin periods are divided into two groups at 20 milliseconds, i.e.,
the millisecond pulsar (MSP) group and others, we find that their
mass averages are, respectively, $M=1.57\pm0.35 \ms$ and
$M=1.37\pm 0.23 \ms$. In the framework of the pulsar recycling
hypothesis, this
 suggests that an accretion of approximately  $\sim0.2\ms$ is sufficient to spin up
 a neutron star and place it in the  millisecond pulsar group.
  Based on these estimates, an  approximate   empirical relation between the accreting
  mass ($\dm$) of recycled pulsar  and its  spin period is proposed
  as $\dm=0.43 (\ms)(P/1 ms)^{-2/3}$.
 If we focus only on the DNS,
 the mass average of all 18 DNSs is $1.32\pm0.14\ms$,  and the mass
averages of the recycled DNSs and the non-recycled NS companions
are, respectively, 1.38$\pm$0.12\ms and 1.25$\pm$0.13 \ms.  This
is consistent with the hypothesis that the masses of both NSs in
DNS system have been affected by accretion.
%
%As a summary, it is inferred  that the mass average of
% isolated NSs without accretion  should be close to  1.4  \ms.
 The  mass average of MSPs is higher than the Chandrasekhar limit
1.44\ms,  which may imply that most of binary MSPs form via the
standard scenario by accretion recycling. If we were to assume
that the mass of a MSP formed by the accretion induced collapse
(AIC) of a white dwarf must be less than 1.35 \ms, then the
portion of the binary MSPs involved in the AICs would  not be
higher than 20\%, which imposes a constraint on the AIC origin of
MSPs.
With  accreting mass from the companion,  the nuclear matter
composition of MSP may experience a  transition from the 'soft'
equation of state (EOS) to a 'stiff' EOS or even neutron to quark
matter. 
 }

\maketitle

\section{Introduction}

% significance of NS mass
Mass is one of the important parameters of a neutron star (NS),
from  which we can  infer the stellar evolution of its progenitor,
 the nuclear matter composition of a compact object (e.g. Haensel et
al. 2007) or its equation of state (EOS), and strength of
gravitational field if the  NS  radius is known.  In other words,
the precise mass measurements can provide  significant tests of
  studies of stellar evolution,  nuclear physics of superdense
matter, and Einstein's  general relativity in the strong gravity
regime (Lattimer \& Prakash 2004, 2007; Kramer \& Stairs 2008), as
well as insight into  binary evolution   since  NS masses are
measured in binary systems.

 A NS is one of the possible ends for a  massive star  with
 mass greater than $\sim$ 4 - 8 \ms. After   having finished burning
 the nuclear fuel, a star undergoes a supernova (SN) explosion, and the central
region of the star collapses under gravity  to form a NS  in the
central  supernova remnant  (SNR) (Haensel et al. 2007).  Hence,
the NS  mass statistics    help  the astronomers to infer the
properties of its progenitor star, and its links to SN and SNR.
However, unlike the other NS parameters, e.g. spin period and
magnetic field, NS mass is only measured in the binary system
(e.g. Freire 2008; Lorimer 2008; Lyne \& Smith 2005). Therefore,
the statistics of the measured NS masses may provide information
 about the  NS accretion history in the binary phases (e.g. Stairs 2004;
Manchester 2004; Bhattacharya \& van den Heuvel 1991).

%mass mearsure

An accurate  measurement of a NS mass  in a pulsar binary system
needs
 five relativistic ※post-Keplerian§ parameters, e.g., the periastron advance, time dilation,
 orbital shrinking rate, and Shapiro delays, which can in principle be measured.
  All of these relativistic parameters place constraints on the NS masses, and when
  three are measured,
   an   accurate determination of NS masses becomes  possible  (e.g. Lorimer 2008;
 Freire  2004, 2008ab, 2009).
The NS masses have been  measured precisely in double neutron-star
(DNS) systems, such as the first  discovered pulsar PSR B1913+16
(Hulse \& Taylor 1975; Taylor \& Weisberg 1982) and double pulsar
system PSR J0737-3039 (Burgay et al. 2004; Lyne et al. 2004;
Kramer \& Stairs 2008),  because   the eccentric orbits of both
systems provide   well-measured relativistic parameters.
 Unlike DNSs,  masses of millisecond pulsars (MSPs)  are not easy  to determine, since their binary  orbits
 are so  circular (or of such low  eccentricity)  that  normally no sufficient
 relativistic effects can be used to
 provide  extra equations to solve the  masses (Freire 2000; Freire et al. 2004).
  Therefore, the masses of MSP systems are often measured with large  errors,
  such as  PSR J0514每4002A (Freire et al.  2004),  except in   cases of  high
eccentricity.
   The  observations of relativistic parameters in pulsar binary systems
   have presented  the first application of general relativity and
   provided  the most widely
   available laboratories for testing theories of gravitation (e.g. Hulse \& Taylor 1975;
   Taylor \&  Weisberg 1982;  Weisberg \&  Taylor 2003;  Thorsett et al. 1993;  Stairs et al. 2002).

Mass measurements are also possible in X-ray binaries, where a
neutron star X-ray pulsar and an optical companion reside. Careful
monitoring of the cyclical Doppler shifts of the pulse period and
Doppler shifts of the spectral features of the optical companion
can be used to determine the orbital period as well as the radial
velocity, which provide/infer the mass function of the system.
Both masses are known when the inclination angle of an eclipsing
binary system can be measured  (e.g. van Kerkwijk et al. 1995;
Jonker et al. 2003).
 The accuracy of this method is not so high as that of measuring DNS mass, usually
  being affected by an error of  about $\sim$ 10\% or more (see  Table 1).

 Thorsett  and Chakrabarty (TC99) presented the results of a statistical study of 19
NS binary systems, and obtained a Gaussian distribution of mass
around  1.35 \ms ,  with a narrow deviation of 0.04 \ms. The
sample has increased significantly since then.  There are now
about 61 NSs with measured (estimated) masses in various types of
binary systems.

%Until now, about 61 NS masses have been measured (estimated) in
%various binary systems, as shown in FIG. \ref{m-list} and  Table
%1-3,  e.g.  the radio pulsars with companions (NS, white dwarf,
%main sequence star), and  X-ray NSs with low or high mass evolved
%stars.  By exploiting these 61 measured NS masses, we investigate
%the statistics and distribution of them and try to update the
%^conclusions what have been achieved by TC99.
%

  In this paper,  we present a statistical analysis of the masses of
NSs in binaries using the current data set, and investigate in
particular the pulsar recycling hypothesis. We present a
compilation of all NS mass observations in Sect. 2. In Sect. 3, we
study the relation between the NS mass and its spin period. Our
conclusions are given in Sect. 4.

\section{Statistics of pulsar masses}

\subsection{NS mass distribution}

 In Tables 1-3,  we   list  all known NSs with measured and
estimated masses, including their binary parameters when
available. In Table 1, we list the 13 systems consisting of X-ray
NSs with low or high mass post-main-sequence star companions. In
Table 2, we  first list the 18 DNSs that have masses with high
accuracies,  then 16 radio pulsars with WD companions, 3 radio
pulsars with the main-sequence star companions  and one uncertain
system. In Table 3, we also list the 10 Galactic radio pulsars
with   WD companions.

\noindent
\begin{figure}
\includegraphics[width=9.0cm, angle=0]{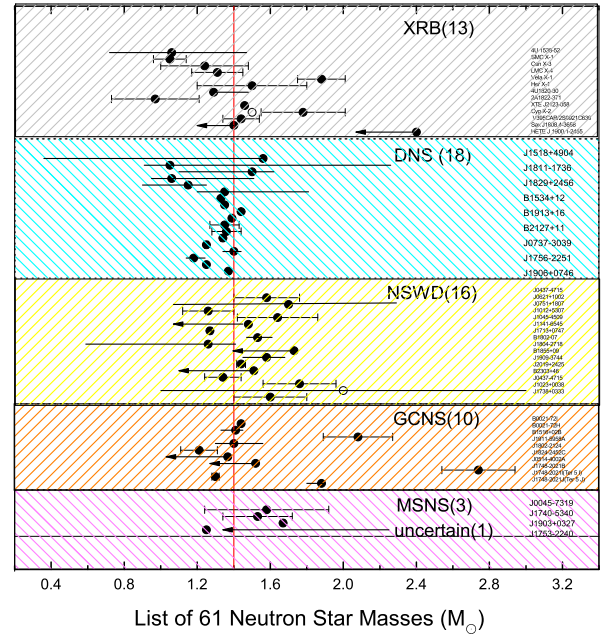}
\caption{List of  61 measured NS masses in the different types of
NS binary systems. Their details and references can be seen in
Table 1-3.  Vertical line M=1.4 \ms { } delineates  the mass mean
value inferred from Gaussian fitting. } \label{m-list}
\end{figure}
%
%\newpage

\begin{figure}
\includegraphics[width=8.0cm,angle=0]{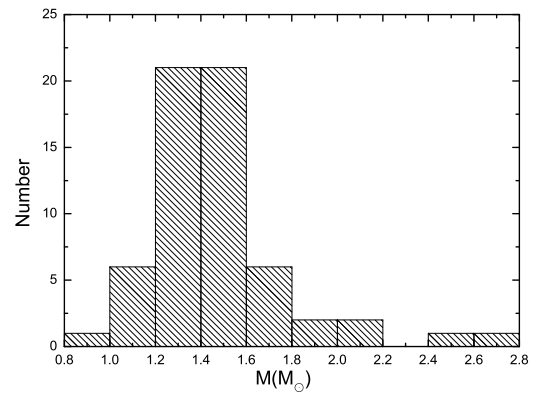}
 \caption{Histogram of 61 measured NS masses.
  A  Gaussian fitting curve is superimposed on the histogram plot,
  with the mass mean value  1.40 \ms { } and standard deviation  0.18  \ms. } \label{m-hist}
\end{figure}

%\vskip 1cm

%All the masses cluster around the value of 1.40 \ms  with the
%standard deviation 0.25\ms,

To illustrate all  NS mass distributions,  a histogram of NS
masses is plotted in Fig.\ref{m-hist},  where a fitted  Gaussian
distribution function is shown  with a mean  mass of  1.40 \ms { }
and a small uncertainty of  0.19 \ms{},  that is slightly higher
than the previous statistical mean value of 1.35 $\pm$ 0.04 {}\ms
by TC99. About $\sim$67\% ($\sim$90\%)  of all NSs are within the
range of 1.2\ms -- 1.6\ms { }(1.0\ms -- 1.8\ms). The NSs with
masses over 1.8\ms { } represent   about $\sim$10\% of all
samples. The maximum and minimum values of NS masses are,
respectively, 2.74 $\pm$ 0.2 \ms {}(J1748-2021B) and 0.97 $\pm$
0.24 \ms {}(2A 1822-371).

It is  interesting to investigate why  the present NS  mass
average is higher than that measured ten years ago. The data of NS
masses by TC99 are based on the DNSs,  which are generally less
than the canonical value of 1.4 \ms.  The present NS mass data
includes all types of binary systems with different evolutionary
histories. In particular, there are many NSWD systems,  which have
significantly high   NS masses as shown in Table 1-3.

It is generally assumed  that MSPs are formed from the spin-up of
a magnetic neutron star caused by  accretion in a binary system
(e.g. Alpar et al. 1982;  Bhattacharya \& van den Heuvel 1991; van
den Heuvel 2004). If the neutron stars were born with the standard
pulsar type fields $\sim 10^{12}$~G, it has to be assumed that the
field decays to $\sim 10^{8-9}$~G by accretion as well. The MSPs
are understood  to be evolutionarily linked  to the long-lived
LMXBs (e.g. van den Heuvel 2004).   The  evidence of a  MSP that
is linked to an   LMXB  was  found with  the discovery of the
first accretion-powered   X-ray pulsar \sax  {} (spin frequency of
401 Hz, Wijnands \& van der Klis 1998).
   %  PSR J0737 - 3039A/B  To test the spin
%
A consequence of the re-cycling hypothesis for the origin of MSPs
is that the mass of a  MSP  should be higher  than that of
non-recycled pulsar. It has long been believed that a MSP should
possess a higher  mass than the canonical value of 1.4 \ms,  e.g.
$\sim 1.8 \ms$, because of the significant amount of accretion
(e.g. van den Heuvel \&  Bitzaraki 1995ab; Burderi et al.  1999;
Stella \& Vietri 1999).
Thus,  if this  relation between MSP mass and accretion  exists,
we may expect to see it in NS mass statistics taken over different
spin period ranges.

\begin{figure}
\includegraphics[width=8.5cm]{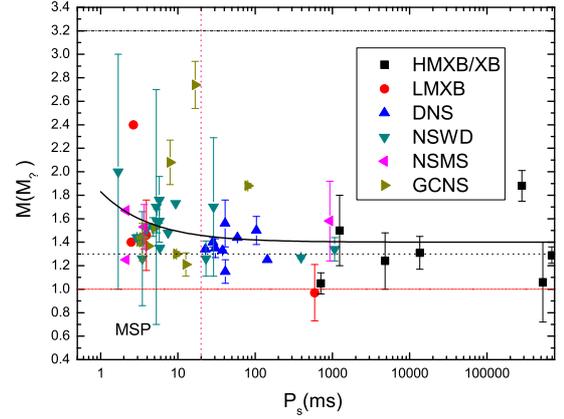}
\caption{Diagram of mass versus spin period for 39 NSs. The
horizontal line M=1 \ms (3.2 \ms) stands for the measured minimum
mass (theoretical maximum mass, see Rhoades  \& Ruffini  1974).
The vertical line at 20 ms separates  the samples into two groups,
MSP ($< 20$ ms) and less recycled NS  ($> 20$ ms). It is found
that the mass averages of two groups are, respectively,
$1.57\pm0.35$ \ms  { } and $1.37\pm0.23$ \ms . The solid curve
stands for the relation between accretion mass and spin period of
recycled pulsar as described in Eq.(1) and (2),
$M=1.40+0.43({P_{s}\ov   {\rm ms}})^{-2/3}$ (\ms) .}
\label{m-spin}
\end{figure}

We first divided all NS samples into two groups, those with spin
periods longer  than and equal  to or shorter  than 20 ms.  The 20
ms dividing line was taken somewhat arbitrarily as the period
below which a pulsar would be classified as a MSP. We find that
the mass averages of MSPs and less recycled NSs are
$1.57\pm0.35\ms$ and $1.37\pm0.23\ms$ , respectively. The expected
trend is therefore clearly seen in the data.
The above trend can also be seen in   Fig. \ref{m-spin}. The mass
systematically decreases with the spin period, or alternatively,
spin-up is associated with an increase in mass of NS.

By dividing the pulsars into three groups, the mass averages are,
respectively,
   M=1.57 $\pm$0.35 \ms ($P<20$ ms),
   M=1.38$\pm$0.23 \ms ($20 ms <P<1000$ ms), and
   M=1.36$\pm$0.24 \ms ($P>1000$ ms).
 Here,  we note  that the average  mass of the recycled pulsar
increases with the stellar spin-up. In general, the spin periods
and magnetic fields (B) of recycled pulsars are  just below the
spin-up line in B-$P_{s}$ diagram of pulsars (e.g. Bhattacharya \&
van den Heuvel 1991; Lorimer 2008),
 where the   B-$P_{s}$  correlation is given by  $P_{s} \sim
 B^{6/7}$ from the accretion-induced magnetic evolution
 model for recycled pulsars (Zhang \& Kojima 2006), the magnetic
 field and accretion mass correlation for recycled pulsars
 is  given approximately by $B \sim \Delta M ^{-7/4}$,  which infers
 a relation  as   $\Delta M  \sim  P_{s}^{-3/2}$.
  On the basis of the  above estimates and arguments,  we  propose an
empirical relation between the accreting mass ($\dm$) of recycled
pulsar and its spin period as

   \be
   \dm = M_{a} (P/{}ms)^{-2/3} \;,
   \ee\label{dm}
 where $M_{a}$ is a characteristic accretion mass when a  pulsar
 is spun-up to one millisecond.
  The mass of recycled pulsar (M)  increases with accretion  and
  is  roughly expressed as,

 \be
 M = M_{0}  +  \dm  \;,
  \ee
where $M_{0}$ is the  mass of NS at birth while NS spin period is
as long as those of HMXBs.

Exploiting Eq.(1) and (2) to fit the NS mass and spin period data
as shown in Fig.3,  we find that  $M_{0} = 1.40\pm0.07$ \ms  { }
and $M_{a} = 0.43\pm0.23$ \ms. Because of the broadness of the
initial NS mass distribution and the large  errors in measuring NS
mass, the fitting COD is as low as 0.07.

%We investigate the fitting function for the M-\pspin, %\be %P = Po - (M-1.4)**2/0.2; M = 1.4 + P/\ee

%\subsection{On DNS mass }

\subsection{Special DNS   mass spectrum}

 The mass average of all eighteen  DNSs in nine systems is $1.32 \pm 0.14 \ms $,
which is systematically lower than that of the  less recycled NS
(M=1.37$\pm$0.23 \ms).
%This phenomenon has been paid attention by
%a couple of researchers before (e.g.  Freire  et al. 2008ab).
%
The mass averages of the nine recycled and non-recycled  DNSs are,
respectively, $1.38\pm0.12\ms$  and $1.25\pm0.13\ms$,  where the
mass of recycled NS is generally higher than that  of non-recycled
one, which may be the indication that either the accretion induces
the  mass increase for the recycled NS or the evolution  of DNS
progenitors makes the mass of non-recycled NS low.
 However, we cannot derive how much mass is accreted into  these
systems, since for two systems (J1811-1736 and J1518+4904) both NS
pair masses have large  differences with large errors, e.g.,  PSR
J1811-1736 with 1.5$^{+ 0.12}_{-0.4}$ \ms {} and 1.06$^{+
0.45}_{-0.1}$ \ms {} (see Table 2).
%
%The  ratio between the recycled NS mass to that of non-recycled
%one is almost unity in most cases, then the longer orbital period
%DNS systems correspond to the higher ratio.

\begin{figure}
\includegraphics[width=8.0cm, angle=0]{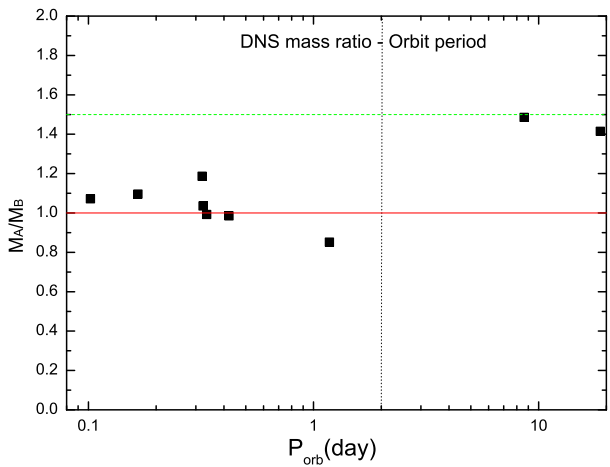}
\caption{Mass ratio versus orbital period diagram for 9 pairs  of
DNSs, where the vertical axis  $M_{\small A}/M_{\small B}$
represents the mass ratio of the recycled NS to non-recycled one.
}
 \label{mab-porb}
\end{figure}

The mass ratios of seven DNSs are close to  unity and those of the
other two with longer orbital periods are higher than  unity, as
shown in Fig.\ref{mab-porb}.  It is too early to draw conclusions
 about any ratio gap, separated by the orbital period  at 2 days,
since fewer  DNS samples are not sufficient to infer a warranty
statistical result.
 The  cause of  the systematically  lower mass values of DNS systems than the typical
 1.4 \ms { } remains unknown.
 We propose  that the evolution  of the DNS progenitors may influence
  or interact  each other, which may be responsible for
   the  particular  mass spectrum distributions shown above.

\noindent

%\subsection{On MSP mass and AIC}

\subsection{On AIC mechanism for MSP formation}

 Although we have focussed on the standard formation model (recycled
NS)  of  MSPs which involves accretion,
 associated  field decay and spin up, other models  are
 possible  (e.g. Kiziltan \& Thorsett 2009abc).
 These  include the often discussed possibility
  of the  accretion induced collapse (AIC) of a
white dwarf onto a neutron star (e.g. van den Heuvel 1994; Verbunt
1990;  Fryer  et al.   1999;  van Paradijs et al. 1997;  Ferrario
\&  Wickramasinghe  2007). In this model, a white dwarf of mass
$>$ 1.2 \ms {} consisting  O, Ne, and Mg (e.g. Nomoto \& Yamaoka
1992) collapses onto a white dwarf because of  the accretion of
matter during the course of binary evolution, where  a NS is
assumed to be born as weakly magnetic and rapidly spinning as
those observed MSPs.

 Hurley et al. (2010) presented a comparative study of
the expected properties of   binary MSPs (BMSPs)    born
 by means of  NS recycling   and  AIC.  They concluded that both
 processes produce
significant populations of BMSPs that  could potentially be
identified with BMSPs.  Furthermore, prior to the detached BMSP
phase at the end of binary evolution, both the NS recycling and
AIC binary systems  may  have experienced  significant phases of
accretion. Nevertheless, the AIC systems are likely on  average to
have accreted less mass.

In Fig.\ref{m-spin}, four  of  twenty-two  BMSPs  have  masses
 of less than 1.35 \ms, which are less than Chandrasekhar mass limit
1.44 \ms, that  may be  candidate  AIC MSPs.  Of course, for a NS
with initial mass of 1.1 \ms, a recycled process will also work by
accreting 0.25 \ms {} from its companion.  If we assume the four
MSPs to be the candidate AICs, then a constraint on  the
production of AIC can be derived that no more than 20\% ($\sim$
4/22) of BMSPs are involved in the AIC processes.
%

% ******

\begin{table*}
 \centering
 \begin{minipage}{175mm}
\caption{Parameters of neutron stars in X-ray binaries}
\label{tab:XB}

\begin{tabular}{cccccccc}
\hline\hline \ System & \ \ \ M($\ms$) \ \ \ & \ \ \ $M_{c}(\ms)$
\ \ \ & \ \ \ $P_{orb}$(d)
 \ \ \ & $P_{spin}$(ms) & \ \ \ eccentricity \ \ \ & $type$ \ \ \ &  \ \  $Refs$ \ \ \\
\hline 4U 1538-52 & $1.06^{+0.41}_{-0.34}$ & $16.4^{+5.2}_{-4.0}$
& 3.73 & $5.28\times10^{5}$ &  0.08 & HMXB & X1 \\ SMC X-1 &
1.05$\pm0.09$ & 15.5$\pm1.5$ & 3.89 & 708 & $<4\times10^{-5}$ &
HMXB & X2 \\ Cen X-3 & 1.24$\pm0.24$ & 19.7$\pm4.3$ & 2.09 & 4814
& $<8\times10^{-4}$ & HMXB & X3 \\ LMC X-4 & 1.31$\pm0.14$ &
15.6$\pm1.8$ & 1.41 & $1.35\times10^{4}$ & $<0.01$ & HMXB & X2 \\
%Vela X-1 & 2.27$\pm0.17$  & 27.9$\pm1.3$  &
%8.96 & $2.83\times10^{5}$ & 0.09 & HMXB & X4 \\
 Vela X-1 &  1.88$\pm0.13$ &  23.1$\pm0.2$ & 8.96 &
$2.83\times10^{5}$ & 0.09 & HMXB & X4 \\
  &  1.86$\pm0.16$ &  23.8$\pm0.2$ & 8.96 & $2.83\times10^{5}$ & 0.09 & HMXB & X4 \\
$4U 1700-37^*$ & 2.44$\pm0.27$ & 58$\pm11$ & 3.41 & No & 0.2 &
HMXB & X5 \\ Her X-1 & 1.5$\pm0.3$ & 2.3$\pm0.3$ & 1.70 & 1240 &
$<3\times10^{-4}$ & XB & X6 \\ 4U1820-30 & $1.29^{+0.19}_{-0.07}$
& $\leq0.106$ & 0.08 & $6.9\times10^{5}$  & No & XB & X7  \\ 2A
1822-371 & 0.97$\pm0.24$ & 0.33$\pm0.05$ & 0.23 & 590 & $<0.03$ &
LMXB & X8 \\ XTE J2123-058 & $1.46^{+0.30}_{-0.39}$ & $0.53^{+0.28}_{-0.39}$ & 0.25 & 3.9 & No & LMXB & X9 \\
Cyg X-2 & 1.78 $\pm0.23$ & 0.60$\pm 0.13$ & 9.84 & No & 0.0 & LMXB & X10 \\
 % & 1.44 $\pm0.06$ & ?  & 9.84 & No & 0.0 & LMXB & X10 \\
  & 1.5  $\pm0.3$  & 0.63 $\pm0.16$   & 9.84 & No & 0.0 & LMXB & X10 \\
V395 CAR/2S 0921每630 & $1.44\pm 0.10$ & $0.35\pm 0.03$ & 9.02 &
No & No & LMXB & X11 \\ Sax J 1808.4-3658& $<$1.4  & $<$0.06  &
0.08 & 2.49 & $<0.0005$ & LMXB & X12 \\ HETE J1900.1-2455 & $<$2.4
& $<0.085$  & 0.06 &
2.65 & $<0.005$ & LMXB & X13 \\
\hline\hline
\end{tabular}

* The compact object may be a black hole (Lattimer \& Prakash 2007).
LMXB---Low-mass X-ray binary, HMXB---High-mass X-ray binary. % %IMXB---Intermediate mass X-ray binary.
  X1---van Kerkwijk et al. 1995 (M, $M_c$, $P_{orb}$, eccentricity); Robba et al.  2001  ($P_s$).
   X2---van Kerkwijk et al.  1995  ($P_{orb}$, eccentricity);  van der Meer et al.
 2005  (M, $M_c$);  van der Meer et al.  2007 ($P_s$).
   X3---van Kerkwijk et al.  1995; Ash et al.  1999 ($P_{orb}$, eccentricity);
van der Meer et al.  2005 (M, $M_c$); van der Meer et al.  2007
($P_s$).
   X4---Quaintrell et al.  2003  (M=2.27,1.88\ms, $M_c$, $P_{orb}$, eccentricity, $P_s$);
    Barziv et al. 2001 (M=1.86\ms).
   X5---Clark et al.  2002  (M, $M_c$); Hammerschlag-Hensberge et al.
 2003 ($P_{orb}$, eccentricity).
   X6---Cheng et al.  1995 ($P_{orb}$, eccentricity); Reynolds et al.  1997  (M, $M_c$);
   Martin et al.  2001 ($P_s$); van der Meer et al. (2007) ($P_s$).
   X7---Wang et al. 2010 (M, $M_c$, eccentricity, $P_s$); Shaposhnikov et al. 2004 (M, $P_{orb}$);  Dib et
   al.  2004 ($P_{orb}$).
   X8---Jonker \&  van der Klis 2001 ($P_{orb}$, eccentricity, $P_s$); Jonker et al. 2003 (M, $M_c$).
   X9---Tomsick et al.  1999 ($P_s$); Tomsick et al.  2002 (M, $M_c$, $P_{orb}$, eccentricity).
   X10---Cowley, Crampton \&  Hutchings 1979 ($P_s$); Orosz \&  Kuulkers  1999 (M, $M_c$, $P_{orb}$,
   eccentricity);
   %  1.44 \ms  Titarchuk \& Shaposhnikov (2002) based on observations of type-I X-ray bursts.
   Elebert,  Callanan \& Torres, et al. {\bf  2009a}.
   X11---Steeghs \& Jonker 1996; 2007 (1.44\ms);  Shahbaz, \& Watson 2007 (1.37㊣0.13 \ms).
 X12 --- Elebert et al. {\bf 2009b}; Chakrabarty  \& Morgan  1998;  Jain, Dutta \&  Paul  ($P_{orb}$).
 X13--- Elebert et al. 2008; Kaaret, Morgan \& Vanderspek  et al. 2006 ($P_{orb}$).

\end{minipage}

\end{table*}
%=================================================

%\newpage
%=================================================

%==================== table 2 ====================
 \begin{table*}
 \centering
 \begin{minipage}{180mm}
\caption{Parameters of radio binary pulsars} \label{tab:RP}
\begin{tabular}{cccccccc}
\hline\hline \ System & \ \ \ M($\ms$) \ \ \ & \ \ \ $M_{c}(\ms)$
\ \ \ & \ \ \ $P_{orb}$(d) \ \ \
 & $P_{spin}$(ms) & \ \ \ eccentricity \ \ \ &  type  \ \ \ &  \ \   Refs  \ \ \\
\hline J1518+4904 & $1.56^{+0.20}_{-1.20}$ & $1.05^{+1.21}_{-0.14}$ & 8.63 & 40.9 & 0.249 & DNS & R1 \\
J1811-1736 & $1.5^{+0.12}_{-0.4}$ & $1.06^{+0.45}_{-0.1}$ & 18.8 &
104.2 & 0.828 & DNS & R2 \\ J1829+2456 & $1.15^{+0.1}_{-0.25}$ &
$1.35^{+0.46}_{-0.15}$ & 1.176 & 41.0 & 0.139 & DNS & R3 \\
B1534+12 & 1.33$\pm0.0020$ & 1.35$\pm0.0020$ & 0.421 & 37.9 &
0.274 & DNS & R4 \\ B1913+16 & 1.44$\pm0.0006$ & 1.39$\pm0.0006$ &
0.323 &
59.0 & 0.617 & DNS & R5 \\ B2127+11C & 1.35$\pm0.080$ & 1.36$\pm0.080$ & 0.335 & 30.5 & 0.681 & DNS & R6 \\
J0737-3039A(B) & 1.34$\pm0.010$ & 1.25$\pm0.010$ & 0.102 & 22.7
(2773) & 0.088 & DNS & R7 \\ J1756-2251 & $1.40^{+0.04}_{-0.06}$ &
$1.18^{+0.06}_{-0.04}$ & 0.320 & 28.5 & 0.181 & DNS & R8 \\
J1906+0746$^{@}$ & 1.25 & 1.37 &
0.166 & 144 & 0.085 & DNS & R9 \\

\hline J0437-4715 & 1.58$\pm0.18$ & 0.24$\pm0.017$ & 5.74 & 5.76 & $1.9\times10^{-5}$ & NSWD & R10 \\
J0621+1002 & $1.70^{+0.59}_{-0.63}(^{+0.32}_{-0.29})$ &
$0.97^{+0.43}_{-0.24}(^{+0.27}_{-0.15})$ & 8.32 & 28.9 & 0.003 &
NSWD & R11 \\
J0751+1807 & $1.26\pm{0.14}$ &
0.19$\pm0.03$ & 0.263 & 3.48 & $3\times10^{-6}$ & NSWD & R12 \\
 & $2.1^{+0.4}_{-0.5}$(corrected) &
0.19$\pm0.03$ & 0.263 & 3.48 & $3\times10^{-6}$ & NSWD & R12 \\
J1012+5307 & 1.7$\pm1.0$  & $0.16\pm 0.02$ & 0.605
& 5.26 & $<10^{-6}$ & NSWD & R13 \\
 &  $1.64\pm 0.22$ & $0.16\pm 0.02$ & 0.605
& 5.26 & $<10^{-6}$ & NSWD & R13 \\
J1045-4509 & $<1.48$ & $~0.13$
& 4.08 & 7.47 & $<10^{-5}$ & NSWD & R14 \\
J1141-6545
& 1.27$\pm0.01$  & 1.02$\pm0.01$ & 0.198 & 394 & 0.172 & NSWD & R15 \\
&
 $1.3\pm 0.02$ & $0.986\pm 0.02$ & 0.198 & 394 & 0.172 & NSWD & R15 \\

J1713+0747 & $1.53^{+0.08}_{-0.06}$($1.6\pm 0.24$) & $0.33\pm
0.04$ & 67.83 & 4.75 & $7.5\times10^{-5}$ & NSWD
& R16 \\ B1802-07 & $1.26^{+0.15}_{-0.67}$ & $0.36^{+0.67}_{-0.15}$ & 2.62 & 23.1 & 0.212 & NSWD & R17 \\
J1804-2718 & $<1.73$ & $~0.2$ & 11.1 & 9.34 & $4\times10^{-5}$ &
NSWD & R18 \\ B1855+09 &
$1.58^{+0.10}_{-0.13}$ & $0.27^{+0.010}_{-0.014}$ & 12.33 & 5.36 & $2.2\times10^{-5}$ & NSWD & R19 \\
J1909-3744 & 1.44$\pm0.024$ & 0.20$\pm0.0022$ & 1.53 & 2.95 &
$~10^{-7}$ & NSWD & R20 \\ J2019+2425 & $<1.51$ & $0.32-0.35$ &
76.5 & 3.93 & $1.1\times10^{-4}$ & NSWD & R21 \\ B2303+46 &
1.34$\pm0.10$ & 1.3$\pm0.10$ & 12.34 & 1066 & 0.658 & NSWD & R22
\\ J0437-4715 & 1.76$\pm0.20$ & 0.25$\pm0.018$ & 5.74 & 5.76 &
$1.918\times10^{-5}$
& NSWD & R23 \\
J1023+0038 & 1.0-3.0 & 0.14-0.42 & 0.198 & 1.69 & $\leq2\times10^{-5}$ & NSWD & R24 \\
%Ter 5 I & $1.3\pm{0.02}$ & 0.24 & 1.328 & 9.57 & 0.428 & NSWD & R25 \\
%Ter 5 J & $1.88^{+0.02}_{-0.08}$ & 0.38 & 1.102 & 80.34 & 0.35 & NSWD & R25 \\
J1738+0333 & $1.6\pm 0.2$ & 0.2 & 0.354 & 5.85 &
$1.1\times{10^{-6}}$ & NSWD & R26 \\ \hline
J0045-7319 & 1.58$\pm0.34$ & 8.8$\pm1.8$  & 51.17 & 926 & 0.808 & NSMS & R27 \\
J1740-5340 & 1.53$\pm0.19$ & $>0.18$ & 1.35 & 3.65 & $<10^{-4}$ & NSMS & R28 \\
% B1259-63 & No & $>3.13$ & 1237 & 47.8 & 0.870 & NSMS & R29 \\
J1903+0327 & $1.67\pm0.01$ & 1.05 & 95.17 & 2.15 & 0.437 & NSMS &
R29 \\ \hline J1753-2240 & $\sim$ 1.25  & $\sim$ 1.25  & 13.64 &
95.1 & 0.304 & uncertain & U
\\ %\hline %B1257+12 & &  &
%66.54 & 6.22 & 0.018 & NS-PC & R29 \\
%B1620-26 & & 0.34$\pm0.04$ & 191.4 & 11.1 & 0.025 & NS-PC & R31 \\

\hline\hline
\end{tabular}
%\end{minipage}{175mm}

%\end{table*}

%\newpage
%\newpage
%\begin{table*}
 %\centering
 %\begin{minipage}{185mm}

%\begin{minipage}{175mm}
  DNS---double neutron star;
NSWD---pulsar-white dwarf binary; NSMS---neutron star/main-sequence
binary;  $^{@}$ The recycled NS should be the companion because of
the strong magnetic field of PSR J1906+0746 $\sim 10^{12}$ G.
R1---Nice et al. 1995 ($P_{orb}$,$P_{s}$, eccentricity); TC99
(M,$M_c$);  Janssen   et al. 2008  ($m_p<1.17$ and $m_c>1.55$\ms).
   R2---Lyne et al. 2001 ($P_{orb}$,$P_{s}$,
eccentricity); Lorimer et al. 2008 (M,$M_c$); Breton, 2009
(M,$M_c$). R3---Champion et al. 2004 ($P_{orb}$,$P_{s}$,
eccentricity); Lorimer et al. 2008 (M,$M_c$); Breton et al. 2009
(M,$M_c$). R4---Wolszczan 1991 ($P_{orb}$,$P_{s}$, eccentricity);
Stairs et al. 2002 (M,$M_c$). R5---Hulse \& Taylor 1975
($P_{orb}$,$P_{s}$, eccentricity); Weisberg \& Taylor 2003
(M,$M_c$). R6---Anderson et al. 1990 ($P_{orb}$,$P_{s}$,
eccentricity); Jacoby, Cameron \&  Jenet et al.  2006; TC99
(M,$M_c$).
% \bibitem{} Jacoby B., Cameron P. \&  Jenet F.  et al.  2006,  ApJ, 644, L113
R7---Burgay et al. 2003 ($P_{orb}$, $P_{s}$, eccentricity); Lyne
et al. 2004 (M,$M_c$). R8---Manchester et al. 2001 ($P_{orb}$,
$P_{s}$, eccentricity); Faulkner et al. 2005 (M,$M_c$).
R9---Lorimer \&  Stairs 2006 ($P_{orb}$, $P_{s}$, eccentricity);
Kasian et al. 2007; Lorimer et al. 2008 (M,$M_c$); Breton et al.
2009 (M,$M_c$). R10---Johnston et al. 1993 ($P_{orb}$, $P_{s}$,
eccentricity); van Straten et al. 2001 (M,$M_c$). R11---Camilo et
al. 1996 ($P_{orb}$, $P_{s}$, eccentricity); Splaver et al. 2002
(M,$M_c$). R12---Lundgren et al. 1995 ($P_{orb}$, $P_{s}$,
eccentricity); Nice et al. 2004 (M,$M_c$); Nice et al. 2005, Nice
et al. 2008 (M,$M_c$). R13---Nicastro et al. 1995 ($P_{orb}$,
$P_{s}$, eccentricity); van Kerkwijk et al. 1996, 2005;
   Callanan et al. 1998; TC99 (M,$M_c$). R14---Bailes et al. 1994 ($P_{orb}$, $P_{s}$,
eccentricity); TC99 (M,$M_c$). R15---Kaspi et al. 2000 ($P_{orb}$,
$P_{s}$, eccentricity); Burgay et al. 2003 (M,$M_c$); Bailes et
al. 2003; Bhat \& Bailes, 2008 (M,$M_c$). R16---Foster et al. 1993
($P_{orb}$, $P_{s}$, eccentricity); Splaver et al. 2005 (M,$M_c$).
R17---D＊Amico et al. 1993 ($P_{orb}$, $P_{s}$, eccentricity);
TC99 (M,$M_c$); Lorimer et al. 2008 (M,$M_c$); Breton et al. 2009
(M,$M_c$); Freire 2000. R18---Lorimer et al. 1996 ($P_{orb}$,
$P_{s}$, eccentricity); TC99 (M,$M_c$); Breton et al. 2009
(M,$M_c$). R19---Segelstein et al. 1986 ($P_{orb}$, $P_{s}$,
eccentricity); Nice,  Splaver \& Stairs 2003 (M,$M_c$).
R20---Jacoby et al. 2003 ($P_{orb}$, $P_{s}$, eccentricity);
Jacoby et al. 2005 (M,$M_c$). R21---Nice et al. 1993 ($P_{orb}$,
$P_{s}$, eccentricity); Nice et al. 2001 (M,$M_c$). R22---Dewey et
al. 1985 ($P_{orb}$, $P_{s}$, eccentricity); Kerkwijk \&  Kulkarni
1999 (M,$M_c$). R23---Johnston et al. 1993 ($P_{orb}$, $P_{s}$,
eccentricity); van Beveren et al. 2008 (M,$M_c$). R24---Archibald
et al. 2009 ($P_{orb}$,$P_{s}$,eccentricity,M,$M_c$).
R26---Jacoby,  PhD thesis, (2004); Freire PhD thesis, 2000.
R27---Bell \& Bessell et al. 1995 (M,$M_c$); Kaspi, Bailes \&
Manchester et al. 1996 ($P_{orb}$, $P_{s}$, eccentricity).
R28---Kaluzny et al. 2003 ($P_{orb}$,M); D"Amico et al. 2001
($P_{s}$,eccentricity,$M_c$). R29---TC99 ($P_{orb}$, $P_{s}$,
eccentricity, $M_c$). R29---Champion et al. 2008 ($P_{orb}$,
$P_{s}$, eccentricity, M, $M_c$); Freire et al. 2009.
%R31---Thorsett, Arzoumanian, Camilo and Lyne 1999; Sigurdsson, %Richer, Hansen,
%Stairs and Thorsett 2003
 U---uncertain companion type, Keith et al. 2009ab  ($P_{orb}$,
$P_{s}$, eccentricity, M, $M_c$).
 \end{minipage}\end{table*}

 %\end{minipage}

%\end{table*}
%=================================================

%==================== table 3 ====================
 \begin{table*}
 \centering
 \begin{minipage}{175mm}
\caption{Parameters of Galactic cluster pulsars} \label{tab:GC}
\begin{tabular}{cccccccc}
\hline\hline \ System & \ \ \ M($\ms$) \ \ \ & \ \ \ $M_{c}(\ms)$ \
\ \ & \ \ \ $P_{orb}$(d) \ \ \ & $P_{spin}$(ms) & \ \ \ eccentricity
\ \ \ & $type$ \ \ \ &  \ \  $Refs$ \ \ \\ \hline
J0024-7204I(B0021-72I)  &
1.44 & 0.15 & 0.23 & 3.49 & $6.3\times10^{-5}$ & GC & G1 \\
 J0024-7204H(B0021-72H) & $1.41^{+0.04}_{-0.08}$ & $0.18^{+0.086}_{-0.016}$ &
2.380 & 3.21 & 0.071 & GC & G1 \\ J1518+0204B(B1516+02B) & 2.08$\pm0.19$ & $>0.13$ & 6.860 & 7.95 & 0.14 & GC & G2 \\
J1911-5958A & $1.40^{+0.16}_{-0.10}$ & 0.18 & 0.837 & 3.48 &
$<10^{-5}$ & GC & G2 \\ J1802-2124 & 1.21$\pm0.1$ & $>0.81$ & 0.699
& 12.65 & $3.2\times10^{-6}$ & GC & G3 \\ J1824-2452C &
$<1.367$ & $>0.26$ & 8.078 & 4.158 & 0.847 & GC & G4 \\
J0514-4002A & $<1.52$  & $>0.96$ & 18.79 & 4.99  & 0.888 & GC & G5
\\ J1748-2021B & $2.74\pm
0.2$  & $>0.11$ & 20.55 & 16.76 & 0.57 & GC & G6 \\ %(Ter5I)J1748-2021I & $<1.96$ & $>0.24$ & 1.328 & 9.57 &
%0.428 & GC & G7 \\ %(Ter5J)J1748-2021J & $<1.96$ & $>0.39$ & 1.102 & 80.34 &  0.350 & GC & G7 \\
J1748-2021I(Ter 5 I) & $1.3\pm{0.02}$ & 0.24 & 1.328 & 9.57 &
0.428 & GC & G7 \\ J1748-2021J(Ter 5 J) &
$1.88^{+0.02}_{-0.08}$ & 0.38 & 1.102 & 80.34 & 0.35 & GC & G7 \\
\hline\hline
\end{tabular}

 GC---Globular cluster pulsars.
 G1---Manchester et al. 1991; Freire et al. 2003;
Lorimer et al. 2008. G2---Wolszczanet al. 1989; Bassa et al. 2006;
Cocozza et al. 2006; Freire et al. 2007; Lorimer et al. 2008;
Freire et al. 2008b. G3---Lorimer et al. 2008; Lorimer et al.
2008; Faulkner et al. 2004; Ferdman et al. 2010
(1.24$\pm0.11$\ms). G4---Ransom and Freire, 2009. G5---Freire \&
Ransom 2007. G6---Freire \& Ransom, 2008; Freire et al. 2008a;
Freire 2009. G7---Ransom et al. 2005.
\end{minipage}
\end{table*}
% on name of J0024-7204I, http://www.johnstonsarchive.net/relativity/binpulstable.html
%=================================================
%\clearpage

\subsection{Pulsar: neutron star   or quark star  ?}

\begin{figure}
\includegraphics[width=8.0cm, angle=0]{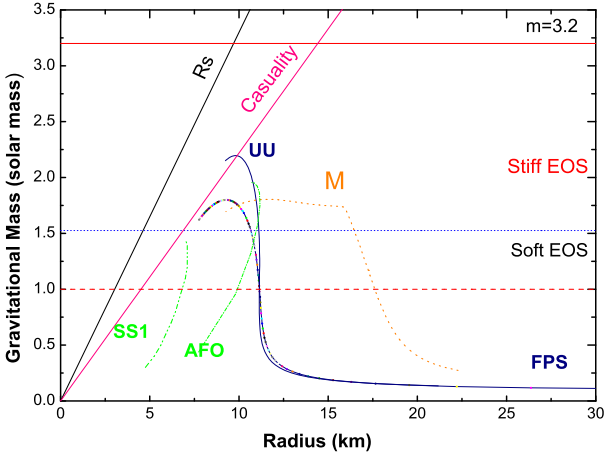}
\caption{NS mass versus radius plot. The EOS curves and straight
lines follow the same meanings as those of Lattimer \& Prakash
(2004,2007) and Miller (2002), where  SS1 and AFO stand for EOSs
of the quark matters. For most NSs with measured masses of 1.0-2.0
\ms, their nuclear matter compositions are difficult to
distinguish  as those of either  neutrons or quarks, since NS
radii cannot be precisely measured in general using  present-day
observations (e.g. Truemper et al. 2004). }
 \label{m-r}
\end{figure}

From the  updated  measured pulsar masses,  we have insufficient
information to clearly infer the nuclear matter compositions
inside the central compact objects,    since we require
measurements of  the stellar radii to determine  the nuclear
matter properties given
  in Fig.\ref{m-r}, a mass-radius plot of compact object.
 Theoretically, pulsars may consist of  hadronic matter
only (Menezes \& Provid\^encia  2004a),  hadronic and quark matter
(hybrid stars)  either bearing or not a mixed phase (Menezes \&
Provid\^encia  2004b; Panda, Menezes \& Provid\^encia 2004; Tatsumi,
Yasuhira \& Voskresensky 2003) or  quark matter only (Menezes,
Provid\^encia \& Melrose {\bf 2006a}; Ivanov et al. 2005). All
calculations depend on choosing  of appropriate equations of state
based on nuclear physics and thermodynamics requirements, which
enter as input to the Tolman-Oppenheimer-Volkoff equations. The
output are a family of stars,   for instance,  with certain
gravitational and baryonic masses, radii,  and central energy. The
maximum gravitational mass and the associated radius are important
constraints on  the equations of state. Generally speaking, the
hadronic matter equation of state (EOS) produces  maximum masses
higher than hybrid stars, which in turn, give slightly higher masses
than quark stars. Radii are usually smaller for quark stars.
However, these results are very model dependent as can easily be
seen from the references mentioned above.

 Therefore,  based on the present results  we cannot determine
 reliably whether   the pulsar is a NS or a quark star
(QS) in this paper. However, we note that the usage of the
terminology NS to denote  the central object of a pulsar is
traditional and  does not imply any  detail of its nuclear matter
composition.

  Theoretically,
   the NS  maximum mass limit of 3.2 \ms {}  was proposed  by Rhoades  \& Ruffini
  (1974). The measured pulsar  masses are then far below this limit, which
    would exclude many known EOS models for the behavior of  matter at supra-nuclear densities.
  The possible existence of high  mass NS  observations favors a stiff EOS (e.g. Ozel 2006; on the NS
stiffness see  Stergioulas 2003).
  The "soft" EOS models predict  lower pressures for a given density, corresponding to a
  less massive star, e.g. $<$ 1.5\ms.
 Recycled NSs in binary systems should find that  the stiffness increases,  and  that the phase
 transition of nuclear matter  may occur (e.g.  Glendenning \& Weber 2001; Menezes  et al. {\bf 2006b}).

The fraction of NSs with masses outside range  1.2 \ms - 1.8 \ms
{}  is less than  20\%,   which  would provide useful information
 about their progenitor properties in most cases.

\noindent

%----------------------------------

%\clearpage

\section{Summary and conclusions}\label{conclusions}

We have studied the statistical distributions of the updated
measurements of  pulsar  masses in binary systems, and the
following conclusions and implications are obtained:

(1) For  61 reliably   measured (estimated) pulsar masses,
 a mass average of M=1.46$\pm$0.3 \ms {} is obtained, which is higher than
 found   (1.35 \ms)  in 1999 by TC99.

(2) Our statistics indicate  that the mass average of the more
rapidly  rotating MSPs (M=1.57$\pm$0.35 \ms  {} for  \pspin $ <
$20 ms) is  higher than that of  the   less  recycled ones
(M=1.37$\pm$0.23 \ms for \pspin $>$ 20 ms).  This implies
 that  the NS masses increase in the accreting spin-up binary systems,  while
  a MSP  accreting  about $\sim$0.2 \ms {}  from its companion appears to be present.
 The relation between the accretion   mass ($\dm$) of recycled pulsar
   and its  spin period is proposed
 to be  $\dm=0.43 (\ms)(P/1{}ms)^{-2/3}$.

(3) The statistics of 18 DNSs indicate  that their
 mass average  M=1.32$\pm$0.14
\ms  is  systematically   lower than the typical mass value
 of the less recycled PSRs,  which seems to imply that the mass formation or
 evolution history of DNS should  differ  from those  of the other binary
 systems.

 (4) Apart from the standard recycled processes for the formation  of MSPs,
 the mechanism by AIC of accreting white dwarfs is investigated by the MSP mass distribution,
 since AIC  needs
 the mass of MSP to be less than the Chandrasehkar mass limit 1.44 \ms. If the AIC explodes
 after accreting  $\sim$ 0.1 \ms  {} of  crust,
  then fewer  than  20 \% of  BMSPs are inferred to be in  the
 AIC processes, which provide a quantitative constraint  on the formation rates of AIC  MSPs.

 (5)
  The  nuclear matter compositions of the less massive DNSs and heavier MSPs may be
  different.  During accretion, the matter phase
   transition from the 'soft' EOS  to 'stiff' EOS,
   or even the matter transition between the neutron and quark  may  be
   possible  (Menezes  {\bf 2006b}), which would provide
    classifications of the nuclear matter  inside DNSs and MSPs.

Moreover, the newly measured mass $ 1.97\pm0.04 $ \ms of a MSP PSR
J1614每2230  with a spin period of 3.15 milliseconds seems to hint
that either MSP accretes more mass from its companion or a high
mass of pulsar is brought in born (Demorest et al. 2010).

\section*{Acknowledgements}
This research has been supported by NSFC (10773017),  NBRPC
(2009CB824800), and CNPq/Brazil.  Thanks are due to the discussions with
 P. C. Freire.  The authors are very grateful for J. Lattimer, C.
 Bassa and G. Janssen  for critic comments which improve the quality of paper.

\end{document}